\documentclass{aastex}
\usepackage{spr-astr-addons}
\usepackage{url}

\RequirePackage{color}

\newcommand{\emaila}{innocent.eya@unn.edu.ng}

\begin{document}

\title{On the pulsar spin frequency derivatives and the glitch activity}
\slugcomment{Not to appear in Nonlearned J., 45.}
%% Running heads
\shorttitle{Pulsar spin frequency derivatives and the glitch activity}
\shortauthors{Eya et al.}

\author{I. O. Eya\altaffilmark{1}} 
\affil{Department of Science Laboratory Technology, University of Nigeria, Nsukka, Nigeria. \\ \emaila{}}

\and \author{J. A. Alhassan\altaffilmark{1}}
\and \author{E. U. Iyida\altaffilmark{1}}
\and
\author{A. E. Chukwude\altaffilmark{1}}
%%\affil{}
\and
\author{J. O. Urama\altaffilmark{1}}
\affil{Department of Physics \& Astronomy, University of Nigeria, Nsukka, Nigeria.}
\email{}

\altaffiltext{1}{Astronomy and Astrophysics Research lab, University of Nigeria, Nsukka, Nigeria.}
%\altaffiltext{2}{Second Alternate Affilation.}
%\altaffiltext{3}{Third Alternate Affilation.}

\begin{abstract}
The number of sudden spin-ups in radio pulsars known as pulsar glitches has increased over the years. 
Though a consensus has not been reached with regards to the actual cause of the phenomenon, the electromagnetic braking torque on the crust quantified via the magnitude of pulsar spin frequency first derivative, $ \dot{\nu} $ is a key factor in mechanisms put across toward the understanding of the underlying principles involved.
The glitch size has been used to establish a quantity used to constrain the mean possible change in pulsar spin frequency $ (\nu) $ per year due to a glitch known as the `glitch activity'.
Traditionally, the glitch activity parameter $ A_{g} $ is calculated from the cumulative glitch sizes in a pulsar at a certain observational time span. 
In this analysis, we test the possibility of 
of quantifying the $ A_{g} $ with the pulsars main spin frequency derivatives (i.e. $ \dot{\nu} $ and $\ddot{\nu} $).
In this approach, the ratio of the frequency derivatives, i.e. $ |\ddot{\nu}|/\dot{\nu}^{2} $ is seen to constrains the glitch activity in radio pulsars. 
The glitch size is found to be independent of the magnitude of the ratio, however, based on the recorded glitch events, the lower end of $ |\ddot{\nu}|/\dot{\nu}^{2} $ distribution appear to have more glitches.
The minimum inter-glitch time interval in the ensemble of pulsars scale with the ratio as $t_{g} \sim 3.35(|\ddot{\nu}|/\dot{\nu}^{2})^{0.23} $.
The $ A_{g} $ quantified in this analysis supports the idea of neutron star inner-crust superfluid being the reservoir of momentum transferred during glitches. It suggests that the moment of inertia of the inner-crust to be at most 10\% of the entire neutron star moment of inertia.
\end{abstract}

% Select between one and six entries from the list of approved keywords.
% Don't make up new ones.
\keywords{pulsars: general --- stars: neutron --- methods: statistical}

%%%%%%%%%%%%%%%%%%%%%%%%%%%%%%%%%%%%%%%%%%%%%%%%%%

%%%%%%%%%%%%%%%%% BODY OF PAPER %%%%%%%%%%%%%%%%%%
\section{Introduction}
\label{sec:intro}

The core remnants of supernovae events are believed to continue their lives as highly magnetized spinning neutron stars; emitting electromagnetic radiations along their magnetic axis.
They are seen as pulsating sources any time their magnetic axis, which is misaligned with the rotational axis sweeps across the earth surface \citep{b8a}, hence the name pulsar. 
On the other hand, there is a class of neutron stars with emission different from the one briefly mention above.
Their emission is believed to originate mostly from the decay of their conspicuous magnetic fields, hence the name magnetar \citep[][and references therein]{b6,Gao2013}. 
They are seen as Anomalous X-ray pulsars (AXPs) and Soft Gamma Repeaters.
%%%%Observationally, they are seen as conventional Radio pulsars, X-ray pulsars, Anomalous X-ray pulsars (AXPs), millisecond pulsars, etc. 
Pulsars spin rates are stable that their precision rivals that of atomic clock, especially in millisecond pulsars \citep{b9}. 
As such any deviation from the predefined spin state is of utmost importance to pulsar researchers. 
Pulsar glitch is one of such deviations. 
In timing data, pulsar glitch is seen as a sudden spin-up, $ \Delta\nu $ in pulsar spin frequency, $\nu$, that is followed by a relaxation phase, which returns the pulsar to a steady state spin frequency \citep [e.g][]{wang00,b7,by}.  
Most of the time, the sudden spin-up is also accompanied by a sudden change in spin frequency derivative, $\Delta\dot{\nu} $.
Formerly, pulsar glitch is seen to be peculiar to radio pulsars.
However, due to increasing timing of other manifestations of neutron stars, glitch events have also been observed in other manifestation of neutron stars, such as magnetars and even in millisecond pulsars \citep[and references therein]{b7}.
As such, glitch events could be a common phenomenon among the pulsar population.

Meanwhile, the origin of glitch or why some pulsars are yet to glitch is still an open debate.
Many physical mechanisms have been put across toward the explanation of possible origin of glitches in pulsars (for review see \cite{b8b}).
Despite that, a comprehensive understanding of pulsar glitch is yet to be attained. 
Most of the mechanisms put across towards the understanding of origin of glitches hinge on glitch sizes and the inter-glitch time intervals and dwell mostly on physics of super-nuclear densities.
For that, pulsar glitch is seen as a viable tool for probing the interior of neutron star, thereby giving insight to the physics of super-nuclear density \citep*[e.g.][]{b53,Chamel12,Chamel13,Wlazlowski,eya17,eya2019,Hujeirat2019,Hujeirat2020}.

A conventional way of explaining the pulsar glitch phenomenon is to parameterize the dynamics of a neutron star as that of two-component system coupled together, but spin differentially \citep{b4,b3,b2,eya17}.
In that scenario, the neutron star components are the observable crust coupled to the core electromagnetically and the inner crust superfluid neutrons.
The superfluid spins via an array of quantized vortices whose area density is proportional to the superfluid angular velocity.
These superfluid vortices are pinned in the inner crust lattice, thereby partially decoupling the inner-crust superfluid from the rest of the other components that co-rotate with the crust.
Inasmuch as the vortices remained pinned, the superfluid velocity remains unchanged.   
Nevertheless, the velocity of the crust keeps changing (decreasing) due to the electromagnetic braking torque on it, making it to spin differentially with respect to the superfluid component.
This process is not expected to continue indefinitely. 
At a point, (a certain right condition, which is not yet well understood, sets in) some of the (or the entire) vortices unpin, migrate outwards transferring their momentum to the crust.
This results in the superfluid spinning down, the crust spinning up: the glitch event \citep{b3}.
The electromagnetic braking torque $ (2\pi I\dot{\nu}) $, which initiate the differential rotation is a function of the stellar moment of inertia $ I $ and the spin frequency derivative, $\dot{\nu}$.
The $\dot{\nu}$ is the only primary variable in the magnitude of the torque, as such; it is the $\dot{\nu}$ that initiates a process that culminates in glitches. 

Another way of viewing the cause of glitches in pulsar is through the mechanism of starquake \citep[e.g][]{b29,b5,Zhou14,eya20b,rencoret2021}.
Though the mechanism was sidelined as a standalone theory for pulsar glitches as it could not accommodate the large and frequent glitches seen in Vela-like pulsars, it is still a viable theory as trigger of glitches \citep[e.g][]{b1,b1a,b7b,eya20b}. 
Models involving starquake view a spinning neutron star as being more oblate than spherical. 
The spherical shape is the supposed equilibrium configuration that the star is to attain, but the crust does not allow a plastic adjustment to that configuration. 
As the neutron star spins down due to electromagnetic braking toque, stress builds up in the crust.
At a critical point at which the crust could no longer sustain the stress on it, a quake occurs, relieving the star of the stress, the oblate is reduced. 
This corresponds to the reduction in stellar moment of inertia.
In this process, conservation of angular momentum entails the spinning up of the pulsar, known as glitch.
Small size glitches ($ \Delta\nu/\nu \lesssim 10^{-9} $) could readily be interpreted using the starquake mechanism.
Indeed, in both glitch mechanisms briefly discussed above, the electromagnetic braking, which is quantified by the magnitude of the spin frequency derivative, 
plays a significant role in the processes.

The glitch size, $ \Delta\nu/\nu $ is in the range $ 10^{-11} - 10^{-5} $ \citep{b7,by}, while the inter-glitch time interval, $ \delta t $ is in the range $ 20 - 10^{4} $d \citep*{b7c}.
Technically if one assumes glitches as a natural phenomenon, which  relives pulsar of stress accumulated over time due to electromagnetic breaking torque, one expects the glitch size to correlate with the inter-glitch time interval, but such is never observed. 
Instead the inter-glitch time intervals for large and small size glitches are seen to be statistically indifferent \citep{melatos2018,b7c,fuentes2019}. 
Establishing a strong correlation of glitch sizes with pulsar spin parameters has been quite challenging.
The glitch size is not a simple function of $ \nu $ or $ \dot{\nu} $ --- that is, the spin parameters, which are directly affected by glitch. 
The only unambiguous correlation so far based on available data is that glitch rate in young and middled-aged\footnote{Age as used here is the assumed spin-down time ($ \tau_{c} = \frac{\nu}{2\dot{\nu}} $) of pulsar known as the characteristic age.} ($ 10^{3} - 10^{5} $ yr) pulsars is higher than that of older ($ \geq 10^{6}  yr$) pulsars \citep[e.g][]{urama99,wang00,b7,eya14,b7fue}.
At present, large glitches are mostly seen in young and middle-aged ($ < 10^{6} yr $) pulsars, as evident in Fig. 1 (Top Panel). 
From Fig. 1, it does appear that the number of large glitches decreases as the pulsar ages and this could be the reason for the prevalence of very small glitches for older pulsars.% may account for the non-observance of glitches 
\begin{figure}[h]
	% To include a figure from a file named example.*
	% Allowable file formats are eps or ps if compiling using latex
	% or pdf, png, jpg if compiling using pdflatex
	\centering
	\includegraphics[width=\columnwidth]{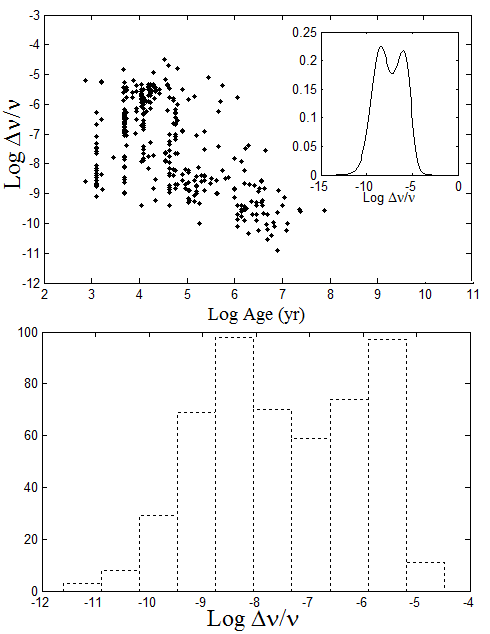}
    \caption{Distribution of glitch sizes ($\Delta\nu/\nu$): top panel is with the pulsar characteristic ages ($\tau_c = \dfrac{\nu}{-2\dot{\nu}}$), while the bottom panel is the histogram. The insert in the top panel is the Kernel smoothing density estimate of the glitch sizes.}
%    \label{fig:example_figure}
\end{figure}

The distribution of glitch sizes is bimodal (Fig.1: bottom panel) as usual \citep[e.g.][]{wang00,b7,eya14,eya17,b7c,eya20a}, with a dip at $ \Delta\nu/\nu \approx 10^{-7} $ hypothetically separating the large glitches from the small ones. 
Nevertheless, to see whether this bimodal is intrinsic to the distribution or due to the number of bins in the histogram, we plotted the Kernel Density Estimate (see, the insert in Fig.1 top panel). 
This also showed a bimodal nature with two distinct peaks at $ \Delta\nu/\nu \approx 10^{-9} $ and $ \Delta\nu/\nu \approx 10^{-6} $ for the small and large glitches respectively and with a dip at $ 10^{-7} $ as the histograms have consistently been.  

On the other hand, the $ \Delta\nu/\nu $ in AXPs are always larger compared to that of other pulsars of similar age. 
Some of their glitches are also associated with radiative changes \citep [e.g.][]{b6,bdk,bka} to the extent that their size could have been enhanced by the radiative change.
Glitch mechanisms that reasonably explain the features of glitches in other manifestation of pulsars is not reasonable enough in AXP glitches.   
Presently, no glitch mechanism has fully incorporated the radiative changes associated with glitches in AXPs. 
In most statistical analysis of neutron star glitches, AXPs are seen as a special case. 
Likewise in this analysis, emphasis is on radio pulsars other than AXPs.

Meanwhile, each glitch event is accompanied by a change in spin frequency and most of the time in frequency derivatives.  
A fraction of these changes are sometimes permanent, while in some events they are transient (if the pulsar recovers back to the pre-glitch spin frequency/frequency derivative during the relaxation phase). 
As such the magnitude of the spin frequency and it's derivative in a given pulsar incorporate how glitches have affected the pulsar over a life time. 
Anyway, a quantity that measures the mean change in pulsar spin frequency per year due to glitch known as Glitch Activity parameter \citep{mckenna90} has been established. 
The glitch activity parameter ($ A_{g} $) is an important parameter in statistical study of glitch events.
Interestingly, it has been show that the glitch activity increases with the magnitude of the spin frequency derivative \citep{mckenna90,urama99,lyne00,b7,b7fue}.

The glitch activity parameter is conventionally calculated from the cumulative glitch sizes in a pulsar over a certain observational period. 
The slope of the cumulative glitch sizes as a function of time from the first glitch readily give the activity parameter \citep* [e.g.][]{b53,b3a}.
Nonetheless, at present, glitch activity can readily be evaluated in pulsars with multiple glitches. 
The activity parameters in such pulsars are highly influenced by the number of large size ($\Delta\nu/\nu \geq 10^{-7} $) glitches in them. 
In that, pulsars with multiple larger glitches have larger activity parameter than the very old and young pulsars known for small size glitches. 
Another issue of interest is even whether a given pulsar has been observed long enough that the current $ A_{g} $ could reflect the actual glitch activity of such a pulsar.  
Therefore, the magnitude of the activity parameter is being biased by the number of recorded glitches in the pulsar.  
Contributions from previous glitches not observed are not accounted for. 
If one is to compare the inter-glitch time intervals with pulsar characteristic ages, ideally one expects a pulsar to have been glitching even before it's discovery. 
The inter-glitch time intervals are size independent and the time preceding large glitches is statistically indifferent with that preceding small size glitches \citep{b7c}. 
Therefore, the inter-glitch time intervals do not bias the magnitude of the activity parameter compared to the size of glitches.
To consider a glitch activity parameter that will not be biased and places all pulsars at equal footing, one needs to consider pulsar spin parameters that have evolved with the pulsar, which is independent of glitch size or observational time span. 
The likely parameters are the spin frequency derivatives, that is, the first ($ \dot{\nu} $) and second ($ \ddot{\nu} $) derivatives.
The present value of $ \dot{\nu} $ incorporates how the $ \nu $ has evolved with time; also it does accommodate how glitches have affected the pulsar over time.
If one could infer the magnitude of the activity parameter from these pulsar spin parameters, it could open a new window for probing the interior of neutron stars.

In this paper, we show how the glitch activity parameters of pulsars can be estimated from the ratio of their spin frequency derivatives. 

\section{Pulsar Spin-down}
The rate of arrival of pulses on a telescope (which tracks the spin frequency, $ \nu $,) after the correction of the effects of interstellar medium and motion of the earth is conventional expressed as a truncated\footnote{other higher terms of frequency derivatives are neglected} Taylor series of the form,\citep{Biryukov2012}
\begin{equation}
\phi(t) = \phi_{0} + \nu(t - t_{0}) + \frac{1}{2}\dot{\nu}(t - t_{0})^{2} + \frac{1}{6}\ddot{\nu}(t - t_{0})^{3} +...+ ,
\label{taylor_series}
\end{equation}
which is expressed in terms of observed spin frequency as
\begin{equation}
\nu(t) = \nu_{0} + \dot{\nu}(t - t_{0}) + \frac{1}{2}\ddot{\nu}(t - t_{0})^{2} + \frac{1}{6}\dddot{\nu}(t - t_{0})^{3} +...+ ,
\label{taylor_series}
\end{equation}
where the rotation phase $ \phi $ and values of the spin frequency and its derivatives are attributed to some reference time $ t_{0} $ \citep{Biryukov2012}.

Long-term observation has shown that pulsars gradually brake as they emit electromagnetic radiation and particle.
The dynamic of the braking is normally expressed as,
\begin{equation}
\dot{\nu} = - K\,\nu^{n}
\label{spindown}
\end{equation}
where n is the braking index and K is a positive constant determined by the mechanism braking the pulsar. 
For braking due to magnetodipole radiation in a vacuum, $n = 3$, for quadrupole magnetic field, $n = 5$, while for pulsar wind braking, $n = 1$. Assuming constancy of K, the braking index is readily ascertain by taking the derivative of Equation ~\ref{spindown}, which gives,
\begin{equation}
n =  \frac{\nu\ddot{\nu}}{\dot{\nu}^{2}}. 
\label{braking_index}
\end{equation}
As such the braking index is just a simple combination of the spin frequency and its first two derivatives.

Concerning the constancy of K, it is worth noting that Equation ~\ref{braking_index} is correct for any braking mechanism \citep{Malov},
likewise, Equation ~ \eqref{FRD},
\begin{equation}
\frac{n}{\nu} = \frac{\ddot{\nu}}{\dot{\nu}^{2}}.
\label{FRD}
\end{equation}
In this analysis, we refers to the right-hand side of Equation ~ \eqref{FRD} as Frequency Derivatives Ratio (FDR). 
Apart from the magnitude of the $ \nu $, which is believed to be a resource from the progenitor star, the derivatives are governed by the mechanism braking the pulsar. 
As such, the derivatives can be used to study the behaviour of pulsars as they spin-down, such as glitches studies.
Glitch parameters have been and are still providing viable tools for probing the interior of neutron stars. 
The $ A_{g} $ is one of such a viable parameter. 
If $ A_{g} $ can be estimated from the FDR, then the $ A_{g} $ from FDR will an alternative parameter for studying the structure of neutron stars without waiting for glitch events.
 In section ~\ref{Glitch Activity and the Spin frequency derivatives}, we show how the $ A_{g} $ obtained from FDR can be used to constrain the neutron star components participating in glitch.

\section{Data Description}
Glitches in this analysis are from the Jodrell Bank Observatory (JBO) pulsar glitch catalogue\footnote{http://www.jb.man.ac.uk/pulsar/glitches.html.}.
The catalogue contains 537 glitches in rotation of 190 pulsars as of the time of this analysis.
Majority of the pulsars have few number of glitches. 
This invariably reduces the number of pulsars available for statistical analysis.
The pulsar spin parameters are from the Australia Telescope National Facility (ATNF) pulsar catalogue\footnote{http://www.atnf.csiro. au/people/pulsar/psrcat} and references therein.
The frequency and its derivatives are long term values.
For the sample for glitch activity analysis, emphasis is on pulsars with multiple glitches of which three of the glitches are large ($ \Delta\nu/\nu > 10^{-7}$), as leaving out the smallest ones (which according to the published distributions are many orders of magnitude smaller) does not result in a strong effect on the activity parameter.
In addition, Crab pulsar (J0534+2200) is included in the sample because it is one of the most widely studied pulsars in terms of pulsar glitch analysis and it has recorded two large glitches.
Pulsars that met these criteria are presented in Table 1 along side with their corresponding frequency first ($ \dot{\nu} $) and second derivatives ($ \ddot{\nu} $), and the characteristic age. 
The sampled pulsars have been observed for decade; as such their long-term spin parameters could be approximated to their precise value. The timing that deduced the spin frequency and its derivatives in each of the pulsars are from the same source, except for PSR J1105 - 6107 in which that of the $\dot{\nu}$ are $\ddot{\nu}$ different \citep[see][]{jbv+19, wang00}. 

From the table, it is readily observed that all the pulsars in the sample are young and middled aged pulsars ($ < 10^{6} yr $).

\begin{table*}[t]
	\centering
	\caption{Glitch and spin parameters of the sampled pulsars with atleast three large glitches.}
	%\label{tab:example_table}
	\begin{tabular}{lcccll} % four columns, alignment for each
		\hline \hline
		Pulsar &$ N_{g} $& $ N_{L} $& $ \tau_{c} $   &$\dot{\nu} $  & $\ddot{\nu} $\\
		J name &&&(kyr)&($10^{-12} s^{-2} $)&($ 10^{-22}$ $ s^{-3}  $)\\
		\hline
		 \hline
0205+6449&13 &7&	5.37      & -44.865(9)$^{a}$	& 58.5153(5)$^{a}$ \\
0534+2200&30 &2&	1.26		& -377.535(2)$^{b} $ 		&111.47(5)$^{b} $  \\
0537-6910&45 &35&	4.93	 &-199.2272(4)$^{c} $		&61(3)$^{c} $\\
0835-4510&20 &18&	11.30   	&-15.666(2)$^{d} $		&10.28(1)$^{d} $ \\
1048-5832&6 &4&	20.30		&-6.298(2)$^{e} $		&1.47(10)$^{e} $ \\
1105-6107&5 &3&	63.30		&-3.963(2)$^{e} $		&-0.54(18)$^{g} $ \\
1119-6127&4 &3&	1.61		&-24.15507(1)$^{f} $		&6.389(4)$^{f} $ \\
1341-6220&23 &16&	12.10		&-6.77115(7)$^{g} $		&-0.1(19)$^{g} $ \\
1413-6141&7 &4&	13.60		&-4.0872(19)$^{h} $		& ---\\
1420-6048&5 &5&	13.00   	&-17.8912(7)$^{i} $		 &---\\
1709-4429&5 &4&	17.50      &-8.857444(12)$^{g} $		&1.731(7)$^{g} $ \\
%%1708-4009& &6&		&	-0.162		&-1.000 \\
1730-3350& 4&3&	26.00   	&-4.3616(9)$^{k} $		&0.619(4)$^{k} $ \\
1731-4744&5 &3&	80.40		&-0.237616(5)$^{g} $		&0.056(13)$^{g} $ \\
1740-3015&36 &11&	20.60		&-1.26557(3)$^{l} $		& ---\\
1801-2451&7 &6&	15.50		&-8.1959(3)$^{l} $		&4.01(10)$^{l} $\\
1801-2304&15 &7&	58.30		&-0.6531075(6)$^{k} $		&0.0112(4)$^{k} $ \\
1803-2137&6 &4&	15.80    	&-7.520039(9)$^{l} $		&2.0793(4)$^{l} $ \\
1826-1334&7 &5&	21.40	 	&-7.3062994(9)$^{l} $		&2.8088(9)$^{l} $ \\
1932+2220&3 &3&	39.80	 	&-2.758388(10)$^{k} $		&0.439(8)$^{k} $ \\
2021+3651&5 &5&	17.20	 	&-8.89419(6)$^{m} $		&10.9(5)$^{m} $ \\
2229+6114&7 &5&	10.50		&-29.37(8)$^{n} $		&---\\
\hline
\end{tabular}\\
\fontsize{8}{10}\selectfont
Note:$ N_{g} $/$ N_{L} $ denotes number of glitches/large glitches. $ \dot{\nu} $ and $ \ddot{\nu} $ are the long term values. The number inside parenthesis is the error at the last digit.
` --- ' denotes unavailability of $ \ddot{\nu} $. Superscripts are the references codes: a -- \cite{lrc+09}, b -- \cite{ljg+15}, c --\cite{mgm+04}, d -- \cite{dml02}, e -- \cite{jbv+19}, f --\cite{wje11}, g -- \cite{wang00}, h -- \cite{kbm+03}, i -- \cite{dkm+01}, k -- \cite{hlk+04}, l -- \cite{ywml10}, m -- \cite{aaa+09d}, n -- \cite{hcg+01}.
\end{table*}

\section{The Glitch Activity}
The glitch activity parameter ($ A_{g} $), which is the mean fractional change in spin frequency of pulsar per year due to glitch is defined as \citep{mckenna90}
\begin{equation}
A_{g} =  \frac{1}{t_{obs}} \sum^{n}_{1} \frac{\Delta\nu_{i}}{\nu},
\label{glitch_activity}
\end{equation}
where $ t_{obs} $, is the observational time span for \textit{n}--number of glitches. 
Where the observational time span is not readily available, $ t_{obs} $ is approximated to be the time span between the first and last glitch in a given pulsar. 
The $ A_{g} $ of the sampled pulsars based on Equation ~\ref{glitch_activity} is presented in Table ~\ref{table_two}, third column. 
Other published values of $ A_{g} $ are presented in the second column.
The $ A_{g} $ has not been constant in magnitude in a given pulsar, it varies from analysis to analysis.
Invariably, this is attributed to changes in $ t_{obs} $ and number of large events involved.

The use of Equation ~\ref{glitch_activity} presumes that a given pulsar is being monitored regularly so that no glitch is missed at the interval, $ t_{obs} $ and the contribution of each glitch to the trend of the cumulative glitch size is even. 
For that, the slope of cumulative glitch size as a function of $ t_{obs}$ conventionally gives the $ A_{g} $.
However, in reality, such regular observation is not obtainable, instead, the pulsars are observed periodically and glitch sizes are not even.
At intervals the pulsar is not observed, glitch(es) could be missed especially the small ones. 
Sometimes in a given pulsar with multiple glitches, one of the glitches could be larger than all other ones put together. 
As such, the $ A_{g} $ obtained in such a situation is strongly biased.
Furthermore, in a non-periodic glitching pulsar, a given observational time may have multiple glitch events, whereas an equivalent one in the past or future may contain no glitch event.
As such, a pulsar seen to have  large $ A_{g} $ in a given epoch may have very low $ A_{g} $ in another epoch. 
Likewise the slope of cumulative glitch size as a function of $ t_{obs}$ will not actually present the $ A_{g} $ especially in non-periodic glitching pulsars and in those of uneven glitch sizes.
In these scenarios, if one is to look back in time and consider the pulsar age compared to the inter-glitch time interval $ t_{g} $, the current $ t_{obs} $ is not long enough to reflect the actual glitch activity in pulsars.
Therefore, Equation (~\ref{glitch_activity}) is just an approximation to the activity parameter based on available data. 
Nonetheless, for a detailed analysis on the uncertainty associated with the use of Equation (~\ref{glitch_activity}) in calculating the glitch activity parameter of pulsars, one can readily see \cite{Montoli2021}.

\begin{table}[]
	\centering
	\caption{Glitch activity of the pulsars.}
	\label{table_two}
	\begin{tabular}{lccc} % four columns, alignment for each
		\hline
		J-name & Ag$^{p}$  & Ag$^{s}$  & Ag$^{f}$\\
			&	($ 10^{-7} yr^{-1}$)& ($ 10^{-7} yr^{-1}$)&	($ 10^{-7}  yr^{-1}$)\\
		\hline	
0205+6449&	&8.04 	&4.15 \\
0534+2200& 0.06$^{a}$,\,\* 0.05$ ^{b} $\,\, &0.17	&0.19 \\
0537-6910&	8.88$^{c,d}$\,\,\, &9.09	&0.34\\
0835-4510&	8$^{a,b}$,  \,\, 7.16$^{c,d}$ \,\, &7.57	&7.30 \\
1048-5832& 	4.02$^{c,d}$\, \,&5.69	&9.36 \\
1105-6107&	&1.65	&3.97\\
1119-6127&	&6.79	&1.79\\
1341-6220&	4$^{b}$,\,\, 6.21$^{c,d}$\,\, &8.56	&0.20\\
1413-6141&	&6.83	& --- \\
1420-6048&	&6.44	& --- \\
1709-4429&	4.02$^{c,d}$\,\, &5.44	&3.00 \\
1730-3350&	&3.91	&3.82\\
1731-4744&	0$^{a}$,\,\, 0.1$^{b}$\,\, &1.45	&0.83 \\
1740-3015&	4$^{a}$,\,\,3$^{b}$,\,\,2.89$^{c}$\,\, &3.08	& ---\\
1801-2451&	5.48$^{c,d}$\,\, &7.11	&9.12 \\
1801-2304&	1$^{b}$,\,\,0.88$^{c,d}$\,\, &0.97	&1.76\\
1803-2137&	5.48$^{c}$ \,\, &7.75	&5.12\\
1826-1334&	5$^{a}$,\,\, 7$^{b}$,\,\,2.92$^{c,d}$\,\, &6.33	&7.65\\
1932+2220&	0$^{a}$,\,\,3.47$^{c}$\,\, &8.16	&6.47\\
2021+3651&	&6.87	&24.22\\
2229+6114&	2.3$^{c}$\,\,&4.15	& ---\\
	\hline
	\end{tabular} \\
	\fontsize{8}{10}\selectfont
Note: Ag$^{p}$, Ag$^{s}$ and Ag$^{f}$ columns are glitch activity calculated from previous works, the glitch sizes and spin frequency derivatives respectively. In Ag$^{p}$ column, the superscripts `a' is for \cite{mckenna90}, `b' for \cite{urama99} and `c' for \cite{b3a} and 'd' for \cite{bho}. `` --- " denotes the values that could not be calculated due to the unavailability of $ \ddot{\nu} $. 
\end{table}

\section{Glitch Activity and the Spin frequency derivatives}
\label{Glitch Activity and the Spin frequency derivatives}
The pulsar spin frequency and it's derivative are seen to be altered in glitches.
We believe that the collative of this alteration is incorporated in the current magnitude of the frequency derivatives.
Meanwhile, the glitch activity is seen to correlate inversely with $ \dot{\nu} $ \citep{lyne00,b7,b7fue}, and $ \dot{\nu} $ is seen to correlate with $ \ddot{\nu} $ \citep{urama06,hobbs2010,Biryukov2012,by}.
Thus a combination of these correlating parameters could be used to estimate the glitch activities of pulsars.

On investigating a possible relationship between the activity parameter and the FDR through a combination plots of the activity parameter with frequency derivatives, a relation of the form $ A_{g} \dot{\nu}^{2} $ $ \sim \ddot{\nu} $ is ascertained.% from a fit on data points.
The relationship is shown in Fig. 2.
The correlation coefficient is 0.92.
In Fig. 2, it is seen that most of the radio pulsars follow a trend defined by the fit on the data points. 
The positions of the most widely discussed pulsars in terms of glitch events (Crab and Vela pulsars) in the plot, are in agreement with the trend. 
\begin{figure}[]
	% To include a figure from a file named example.*
	% Allowable file formats are eps or ps if compiling using latex
	% or pdf, png, jpg if compiling using pdflatex
	\centering
	\includegraphics[width=\columnwidth]{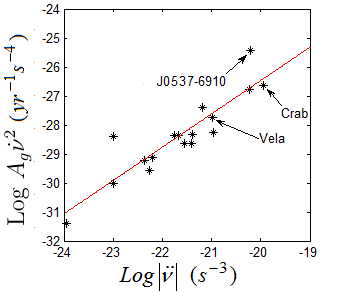}
    \caption{$ A_{g} \dot{\nu}^{2} $ as a function of $ \ddot{\nu} $. This plots contains only the pulsars in our sample that have measured values of $ \ddot{\nu} $.}
    %\label{fig:example_figure}
\end{figure}
 
From Fig. 2, equation of the form, 
\begin{equation}
Log\, (A_{g}\dot{\nu}^{2}) = 1.14 \,Log\, |\ddot{\nu}| +\,Log\,(1.51 \times 10^{-4}),
\end{equation}
is readily obtained.
This presents an alternative way of calculating the glitch activity in pulsars (in unit of $ 10^{-7} yr^{-1}$) as
\begin{equation}
A_{g} \approxeq \frac{|\ddot{\nu}|^{m}C}{\dot{\nu}^{2}}
\label{activity_FDR}
\end{equation}
where m and C are 1.14 and 1.51 $\times 10^{-4} $ respectively.
Equation ~\ref{activity_FDR} shows that with the pulsar spin frequency derivatives [i.e. $ \dot{\nu} $ and $ \ddot{\nu} $] , one can readily estimate it's activity parameter. 
With Equation ~\ref{activity_FDR}, the glitch activity parameter of the pulsars were calculated and the values presented in Table 2 fourth column.  

\begin{figure}
	% To include a figure from a file named example.*
	% Allowable file formats are eps or ps if compiling using latex
	% or pdf, png, jpg if compiling using pdflatex
	\includegraphics[width=\columnwidth]{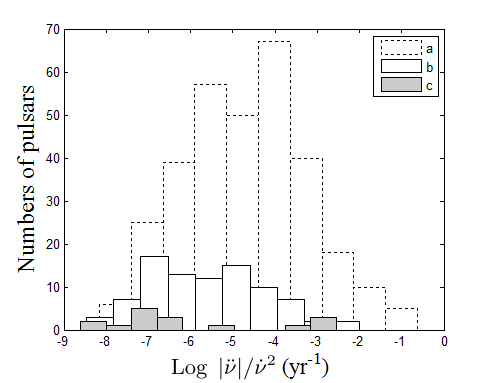}
	\centering
    \caption{Distribution of the ratio $ |\ddot{\nu}|/\dot{\nu}^{2} $ in ensemble of pulsars. Histogram `a' is for pulsars without a glitch, while histogram b and c are for pulsars with less than three large glitches and those with three or more large glitches respectively. Note: This distribution contains only pulsars that have measured values of $ \ddot{\nu} $.}
    %\label{fig:example_figure}
\end{figure}

\begin{figure}
	% To include a figure from a file named example.*
	% Allowable file formats are eps or ps if compiling using latex
	% or pdf, png, jpg if compiling using pdflatex
	\includegraphics[width=\columnwidth]{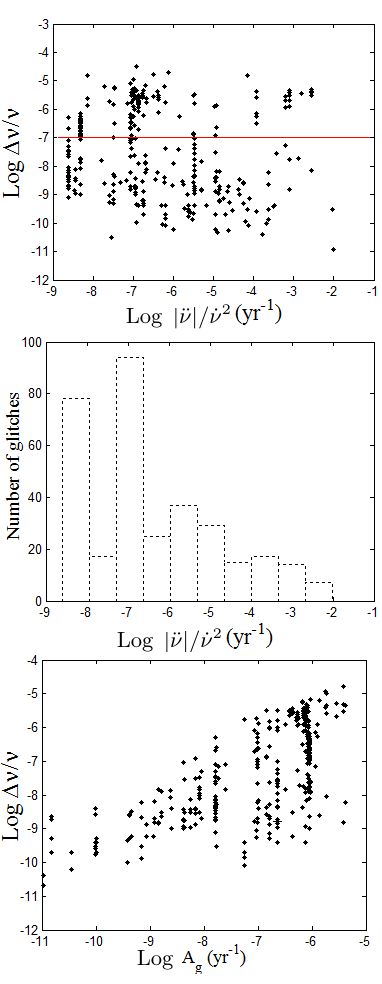}
	\centering
    \caption{ Distribution of glitches with $ |\ddot{\nu}|/\dot{\nu}^{2} $ and $ A_{g} $. Top panel is with glitch sizes, middle panel is with number of events, while bottom panel is with glitch activity parameter.  Note: This plot contains only pulsars that have measured values of $ \ddot{\nu} $. The horizontal line at the top panel is hypothetical demarcation of large glitches from small glitches corresponding to the dip in Fig. 1 bottom panel.}
    \label{FGS_and_FDR}
    %\label{fig:example_figure}
\end{figure}

Concerning Equation ~\ref{activity_FDR} the glitch activity is seen to be constrained by the frequency derivative ratio, $|\ddot{\nu}|/\dot{\nu}^{2}$. 
To ascertain how pulsars with glitches and those without glitches are distributed in the plane of $|\ddot{\nu}|/\dot{\nu}^{2}$, a histogram plot of the distribution of $ |\ddot{\nu}|/\dot{\nu}^{2} $ is analysed.
The result is shown in Fig. 3.
The difference in the heights of the bins in the three histogram is just as a result of the number of pulsars involved in each group.
From the plot, it is readily observed that the range of the ratio do not differ significantly in pulsars with glitches and in those without glitch. 
The range of the distribution in each of the groups (i.e. a, b, c,) are similar.
However, it is seen that no pulsar of $ |\ddot{\nu}|/\dot{\nu}^{2} \lesssim 10^{-8}$ $yr^{-1}$ is yet to glitch and there is no pulsar of $ |\ddot{\nu}|/\dot{\nu}^{2} > 10^{-2}$ $yr^{-1}$ that have glitched, though only few number of pulsars are in these categories.
Furthermore as $ |\ddot{\nu}|/\dot{\nu}^{2} $ constrains the $ A_{g} $, which is conventionally evaluated from, $ \Delta\nu/\nu $, we investigate the possible relationship between $ |\ddot{\nu}|/\dot{\nu}^{2} $ and $ \Delta\nu/\nu $. 
To do that, a plot of $ \Delta\nu/\nu $ as function of $ |\ddot{\nu}|/\dot{\nu}^{2} $ and $ A_{g} $ are made.
The result is shown in Fig. 4.
Interestingly, current glitch sizes ($10^{-11}  \leq \Delta\nu/\nu \leq 10^{-5}$) are feasible across the range defined by the magnitudes of $ |\ddot{\nu}|/\dot{\nu}^{2} $ in the ensemble of pulsars with recorded glitch (Fig. 4, Top Panel).
This is an indication that the magnitude of the quantity, $ |\ddot{\nu}|/\dot{\nu}^{2} $, does not influence the size of glitches. 
As such, using the spin frequency derivatives to calculate the glitch activity parameter, is not biased by number of recorded large glitches in a pulsar like the traditional method (i.e. using Equation ~\ref{glitch_activity}, see bottom panel of Fig. ~\ref{FGS_and_FDR}).
This is understandable as glitch activity is just the mean of possible change in pulsar spin frequency in a pulsar in a given period due to glitch. 
Nonetheless, there are more of large glitches at the left-hand-side of the distribution, while the concentration of large glitches is centred around $ |\ddot{\nu}|/\dot{\nu}^{2} \sim 10^{-7}$ $yr^{-1}$, likewise there are still substantial number of small sized glitches at that region. 
Meanwhile, a peculiar feature in the distribution is that, the number of events decrease with the increasing magnitude of $ |\ddot{\nu}|/\dot{\nu}^{2}$. 
This is conspicuously seen at the bottom panel of Fig. 4. 
This is just an indication that slow braking pulsars glitch less frequently.
The distribution peaks at $ |\ddot{\nu}|/\dot{\nu}^{2} = 10^{-7} (yr^{-1})$, indicating the region of highest concentration of recorded glitch events. 
Interestingly, this is the unity of which the glitch activity parameter is measured. 
The first three bins contain 57\% of glitches in the distribution likewise these bins contain the pulsars, which have the the highest number of events per pulsar and glitch rate.   

\begin{figure}[h]
	% To include a figure from a file named example.*
	% Allowable file formats are eps or ps if compiling using latex
	% or pdf, png, jpg if compiling using pdflatex
	\includegraphics[width=\columnwidth]{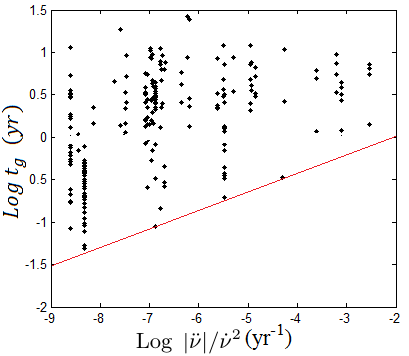}
	\centering
    \caption{A plot of inter-glitch time interval versus FDR. The fit indicates how the minimum inter-glitch time interval scales with the FDR.}
%    \label{fig:example_figure}
\end{figure}

On the other hand, there is wide range of inter-glitch time intervals across the distribution of $ |\ddot{\nu}|/\dot{\nu}^{2}$ as shown in Fig. (5).
The minimum inter-glitch time intervals (in unit of years) scale as $t_{g} \sim 3.35 (\vert\ddot{\nu}\vert/\dot{\nu}^{2})^{0.23}  $, while the maximum $ t_{g}$ appears to be similar across the range of the distribution.
This just indicates that fast braking pulsars are more likely to have a shorter inter-glitch interval than a slow one. As such, objects that have a high spin-down rate like the Crab and J0537 - 6910 have minimum inter-glitch time intervals compared to others. From the scaling, the Crab pulsar, and J0537-6910 have it to be 13 and 15 days respectively, while moderate braking objects like the Vela pulsar have it to be 32 days. 
If compared with available data, that of Crab is 30 days, that PSR J0537-6910 is 18 days, while what is observed for the Vela pulsar is 32 days. 
Thus the scaling relation is in line with observation. 
In addition, this scaling relation just justifies why there is a preponderance of glitches in pulsars of low FDR as evidently seen in Figure 4.

The $ A_{g} $ is a vital tool in constraining the magnitude of neutron star components participating in glitch through the Fractional Moment of Inertia [i.e. the ratio of inner-crust superfluid moment of inertia ($ I_{s} $) to that of the entire neutron star ($ I $)] \citep{b53,b3a,eya17,eya2019}.
In what follows, we constrain the FMI using FDR.

\begin{figure}[t]
	% To include a figure from a file named example.*
	% Allowable file formats are eps or ps if compiling using latex
	% or pdf, png, jpg if compiling using pdflatex
	\centering
	\includegraphics[width=\columnwidth]{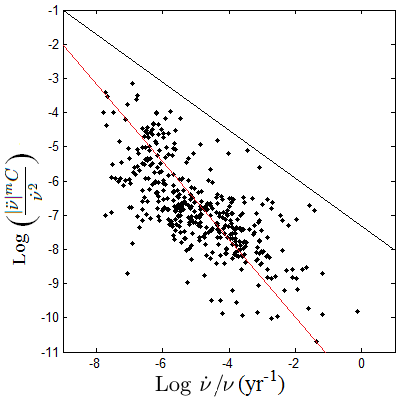}
    \caption{Estimated glitch activity parameter versus $ \dot{\nu}/\nu $ in ensemble of radio pulsars. The lower line (red line) is the line of best fit, while the upper line (black line) is just a hypothetical upper band.}
    \label{Ag_nudot_nu}
%    \label{fig:example_figure}
\end{figure}

Following \cite{eya17}, who calculated FMI using the frequency first derivative as one of the input parameters, we have;
\begin{equation}
\frac{I_{s}}{I} = -\sum\dfrac{1}{\dot{\nu}}\frac{\Delta\nu}{t_{g}}.
\label{equ_FMI}
\end{equation}
Using Equation ~\ref{activity_FDR} and ~\ref{equ_FMI}, gives
\begin{equation}
A_{g}\cdot \dfrac{I_{s}}{I} = - \frac{|\ddot{\nu}|^{m}C}{\dot{\nu}^{2}}\cdot \sum\dfrac{1}{\dot{\nu}}\frac{\Delta\nu}{t_{g}},
\end{equation}
which when combined with Equation ~\ref{glitch_activity} reduces to 
\begin{equation}
\frac{|\ddot{\nu}|^{m}C}{\dot{\nu}^{2}} = - \dfrac{I_{s}}{I}\frac{\dot{\nu}}{\nu}.
\label{FMI_constrian}
\end{equation}
With Equation ~\ref{FMI_constrian}, the FMI of the sampled pulsars are obtained (Table ~\ref{table_three} Column 4).
To average over ensemble of pulsars, a plot of Equation ~\ref{FMI_constrian} is made (Figure ~\ref{Ag_nudot_nu}). 
The logarithm plot of Equation ~\ref{FMI_constrian} gives the average value of FMI for the radio pulsars as the intercept on the vertical axis.
From the red line (lower line), $ I_{s} \approx 10^{-2}I $, which is in favour of the notion that the magnitude of neutron star inner-crust superfluid moment of inertia participating in glitch is just a few percent of that of the entire star \citep[e.g., see:][]{b53,eya17,b7fue}. 
This result shows that the activity parameter estimated from the spin FDR is as good as that estimated from the glitch sizes.
The black line (upper line), which is fitted by an eye-view is a hypothetical upper-band for $ I_{s} $.
Interestingly, this line gives an upper-limit of $ I_{s} \approx 10^{-1}I $. 
This limit is in line with the findings of \cite{steiner2015} who shows that, if one explore the uncertainties in equations-of-state determining neutron star structure that the inner-crust can be as large as 10\% of the entire neutron star. With this, we suggest that there could be a connection between the inner-crust superfluid moment of inertia and the breaking index, n (see Equation ~\ref{braking_index} concerning Equation ~\ref{FMI_constrian}).

\begin{table}[h]
	\centering
	\caption{Glitch activity parameters of the pulsars with the Fractional Moment of inertia.}
	\label{table_three}
	\begin{tabular}{lccc} % four columns, alignment for each
		\hline
		J-name & Ag$^{f}$  & $\dot{\nu}/\nu$  & FMI\\
			&	($ 10^{-7} yr^{-1}$)& ($ 10^{-5} yr^{-1}$)&	($ 10^{-2}$)\\
		\hline	
0205+6449	&	4.15	&	9.30	&	0.446	\\
0534+2200	&	0.19	&	39.78	&	0.005	\\
0537-6910	&	0.34	&	10.14	&	0.034	\\
0835-4510	&	7.29	&	4.42	&	1.651	\\
1048-5832	&	9.36	&	2.46	&	3.808	\\
1105-6107	&	3.96	&	0.79	&	5.012	\\
1119-6127	&	1.79	&	31.10	&	0.058	\\
1341-6220	&	0.20	&	4.13	&	0.048	\\
1709-4429	&	2.99	&	2.86	&	1.045	\\
1730-3350	&	3.82	&	1.92	&	1.991	\\
1731-4744	&	8.32	&	0.62	&	13.37	\\
1801-2304	&	1.76	&	0.86	&	2.052	\\
1801-2451	&	9.11	&	3.23	&	2.819	\\
1803-2137	&	5.12	&	3.17	&	1.613	\\
1826-1334	&	7.64	&	2.34	&	3.264	\\
1932+2220	&	6.46	&	1.26	&	5.135	\\
2021+3651	&	24.18	&	2.91	&	8.304	\\
\hline

	\end{tabular} \\
 
\end{table}

\section{Conclusions}

The $ A_{g} $ estimates the overall fraction of the spin-down that has been reversed by glitches over the time the pulsar was observed.
In this analysis, we have demonstrated that the $ A_{g} $ can be estimated from the spin-down rate of pulsars via $ |\ddot{\nu}|/\dot{\nu}^{2}$. 
This method is seen not to be biased by the size of glitches in a given pulsar, instead, it is based on the current magnitude of the frequency derivatives. 

The results of this analysis is in line with recent analysis, which suggests that a parameter strongly tied with time, plausibly the spin frequency derivative, plays a similar role for both large and small size glitches as $ t_{g} $ is size independent \citep{b7c}.
Based on available glitch mechanisms it is believed that electromagnetic braking torque on the neutron star crust drives pulsar glitches. 
Without it, the neutron star components will be co-rotating and there will be no glitches.
The electromagnetic braking torque, which initiates a process that culminates in glitches is a function of the $ \dot{\nu} $.
Therefore, $ |\ddot{\nu}|/\dot{\nu}^{2}$ can be a measure of glitch activity.

On average, radio pulsars of large frequency derivative (such as the Crab pulsar), are known for small size glitches ($ \Delta\nu/\nu \approx 10^{-9} $), whereas objects like Vela pulsar ($ \Delta\nu/\nu \approx 10^{-6} $) are known for large size glitches.
Meanwhile, this notion is gradually being challenged with recent observation of a large glitch ($ \Delta\nu/\nu > 10^{-7}$) in Crab pulsar \citep{Shaw2018}. % and a very small glitch in Vela pulsar \citep{Jankowski2015}.  
 Though with the outcome of the distribution of $ |\ddot{\nu}|/\dot{\nu}^{2}$ with respect to glitch sizes (Fig. 4), it is seen that such a large glitch is feasible for the Crab pulsar likewise for any other glitching radio pulsar of which $ |\ddot{\nu}|/\dot{\nu}^{2} < 10^{-2} yr^{-1}$. 
What may differ is the frequency of such events. 
Apart from PSR J$1740-3015$ that is well known for mixture of glitch sizes, many pulsars are becoming candidate of mixed glitch sizes, which is an actually in line with the distribution of glitch sizes in $ |\ddot{\nu}|/\dot{\nu}^{2}$ plane (Fig. 4. top panel).
With more glitches becoming available, this picture is expected to be clearer. 

The $ A_{g} $ has not been constant in magnitude in a given pulsar, it varies from analysis to analysis.
Invariably, this is attributed to changes in $ t_{obs} $ and number of large events involved. 
Nonetheless, with this approach via frequency derivatives, the magnitude of $ A_{g} $ in a given pulsar could be fairly constant.

On the other hand, accurate measurement of $ \ddot{\nu} $ is still a serious challenge in pulsar timing.
To this moment, they have been determined precisely enough in a few known pulsars.
The spin frequency and its derivatives can be measured during long-term observation. 
We believe that long-term values of $ \ddot{\nu} $ are good enough and hopeful tool for this kind of studies. 
This is because pulsar timing has a clear and systematically laid down technique. 
The sources of uncertainties anticipated at the time of arrival of pulses in the telescope are distinguished and removed during data reduction. 
As such, there is no permissible reason to believe that the extensive cubic trends, undoubtedly seen in phase solution for hundreds of pulsars \citep{hlk+04}, are artifacts \citep{Biryukov2012}. 
In that, all correctly measured $ \ddot{\nu} $ values are certainly owing to the uniqueness of pulsar spin-down. 
Thus one can use them to study phenomena associated with pulsar spin-down.
Glitches are one of the phenomena encounter in spin-down evolution of pulsars. 
The glitch activity estimated in this analysis concurred to the notion that a few percent of the stellar moment of inertia resides in a region housing the momentum transferred in glitches. 
Thereby giving credibility to the long-term values of the frequency derivatives used in this analysis.

In conclusion, we have shown that with pulsar main spin frequency derivatives, its glitch activity can be estimated and our result is consistent with observation. 
$ |\ddot{\nu}|/\dot{\nu}^{2} $ could be a likely measure of activity parameters in pulsars.

\section*{Acknowledgements}
We wish to thank the anonymous reviewer whose suggestions led to a salient result unimagined to us. That is the possibility of constraining the FMI with FDR. We are grateful to you.
%%colleagues, acknowledge funding agencies, telescopes and facilities used etc.
%Try to keep it short.
\section{Statements and Declarations}
\subsection{Author Contribution}
I. O. Eya is the Principal Investigator and the corresponding author. He and other authors contributed in the interpretation of the results and in writing writing of the paper.
\subsection{Funding information}
There is no funder to this work. 
\subsection{Data availability}
The pulsar glitch data is from \citep{b7} and the Jodrell Bank Observatory (JBO) pulsar glitch catalogue, which is readily available at http://www.jb.man. \\ac.uk/pulsar/glitches.html and references therein.
The pulsar spin parameters are from the Australia Telescope National Facility (ATNF) pulsar catalogue, which is readily available at http://www.atnf.csiro.au/people/pulsar/psrcat and references therein..
\subsection{Conflict of interest}
The authors declare that they have no conflicts of interest.

%%%%%%%%%%%%%%%%%%%%%%%%%%%%%%%%%%%%%%%%%%%%%%%%%%

%%%%%%%%%%%%\begin{flushleft}

%%%%%%%%%%%%\end{flushleft}%%%%%%%% REFERENCES %%%%%%%%%%%%%%%%%%

% The best way to enter references is to use BibTeX:

%\bibliographystyle{mnras}
%\bibliography{example} % if your bibtex file is called example.bib

% Alternatively you could enter them by hand, like this:
% This method is tedious and prone to error if you have lots of references

%%%%%%%%%%%%%%%%%%%%%%%%%%%%%%%%%%%%%%%%%%%%%%%%%%

% Don't change these lines

\label{lastpage}
\end{document}